\documentclass{ws-ijmpe}
\renewcommand{\vec}[1]{\mbox{\boldmath $#1$}}

\begin{document}

\markboth{K. Hagino,H. Sagawa, and P. Schuck}
{Di-neutron correlation in light neutron-rich nuclei}

\catchline{}{}{}{}{}

\title{Di-neutron correlation in light neutron-rich nuclei}

\author{K. Hagino$^1$, H. Sagawa$^2$, and P. Schuck$^{3,4}$}
\address{$^1$
Department of Physics, Tohoku University, Sendai, 980-8578,  Japan \\
$^2$
Center for Mathematical Sciences,  University of Aizu,
Aizu-Wakamatsu, Fukushima 965-8560,  Japan \\
$^3$
Institut de Physique Nucl\'eaire, CNRS, UMR8608, Orsay, F-91406, 
France \\
$^4$
Universit\'e Paris-Sud, Orsay, F-91505, France}

\maketitle

\begin{history}
\received{\today}
\end{history}

\begin{abstract}
Using a three-body model with density-dependent 
contact interaction, 
we discuss the root mean square distance between 
the two valence neutrons 
in $^{11}$Li nuclues 
as a function of the center of mass of the neutrons 
relative to the core nucleus $^9$Li. 
We show that 
the mean distance takes a pronounced minimum 
around the surface of the nucleus, 
indicating 
a strong surface 
di-neutron correlation. We demonstrate that 
the pairing correlation plays an essential role in this behavior. 
We also discuss the di-neutron structure in the $^8$He nucleus. 
\end{abstract}

\section{Introduction}

The idea of di-neutron correlation can date back to the Migdal's paper in 
1972\cite{M73}, in which he argued that two neutrons may be bound in finite 
nuclei even if they are not bound in the vacuum. This idea was explicitly 
exploited by Hansen and Jonson \cite{HJ87}, where they proposed the 
di-neutron cluster model and sucessfully analysed the matter radius 
of $^{11}$Li. They also predicted a large Coulomb dissociation 
cross section of the $^{11}$Li nucleus. 
Although the di-neutron correlation in 
the $^{11}$Li nucleus has been discussed for two decades since 
the publication of Hansen and Jonson, 
it is only recently that
a strong indication of
its existence has been obtained experimentally in the
Coulomb dissociation of $^{11}$Li \cite{N06}.
The new measurement has stimulated lots of theoretical
discussions on the di-neutron correlation, not only in
the 2$n$ halo nuclei, $^{11}$Li and $^6$He
\cite{BBBCV01,HS05,HSCP07,BH07},
but also
in medium-heavy neutron-rich nuclei \cite{HTS08,MMS05,PSS07}
as well as in infinite neutron matter \cite{M06,MSH07}.

In this contribution, we present our recent activities on 
the di-neutron correlation in light neutron-rich nuclei, forcusing 
especially on the size of Cooper pair in the $^{11}$Li nucleus. 
We also discuss the di-neutron correlation in a 4$n$ nucleus, $^8$He. 

\section{Surface di-neutron correlation in $^{11}$Li}

In Ref. \cite{HSCP07}, 
we have 
used a three-body model with density-dependent contact 
interaction \cite{BE91,EBH99} to study 
the two-neutron 
Cooper pair in $^{11}$Li 
at various positions $R$ from the center to the surface of the nucleus 
(this calculation is essentially equivalent to the particle-particle 
Tamm-Dancoff approximation \cite{RS80,PVM02}). 
We have found that i) the two-neutron wave function 
oscillates near the center whereas it becomes similar to 
that for a bound state around the nuclear surface,
and ii) the local pair coherence length has a
well pronounced minimum around the nuclear surface. 
This is
qualitatively the same behavior as found in neutron matter \cite{M06}, 
and has subsequently been confirmed in heavier
superfluid nuclei as well \cite{PSS07}. 

\begin{figure}[htb]
\centerline{\psfig{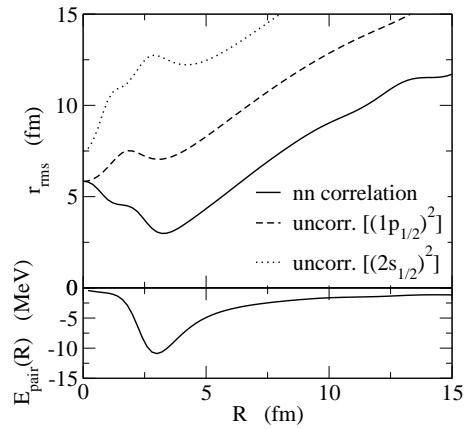}}
\caption{
The root mean square distance $r_{\rm rms}$ for the neutron Cooper pair in
$^{11}$Li as a function of the nuclear radius $R$. 
The solid line shows the result of three-body model calculation 
with density-dependent contact pairing force, 
while the dashed and the dotted lines are obtained 
by switching off the neutron-neutron interaction and assuming the 
[(1p$_{1/2})^2$] and [(2s$_{1/2})^2$] configurations,
respectively. 
The local expectation value of the neutron-neutron interaction
is also shown in the lower 
panel. 
}
\end{figure}

It is important to notice here that 
the pairing interaction changes the 
structure of the Cooper pair {\it strongly} and {\it qualitatively}. 
To demonstrate this, we show in Fig. 1 the local 
coherence length of the Cooper pair in $^{11}$Li
as a function of 
the nuclear radius $R$ obtained with and without the 
neutron-neutron ($nn$) interaction. 
For the uncorrelated calculations, we consider both the
[(1p$_{1/2})^2$] and [(2s$_{1/2})^2$] configurations. 
One can see that, in the non-interacting case, 
the Cooper pair 
continuously 
expands, as it gets farther away from the center of the nucleus. 
In marked contrast, in the interacting case it becomes {\it smaller} going from 
inside to the surface before expanding again into the free space 
configuration. This is nothing more than the 
pronounced strong coupling, 
{\it i.e.,} the BEC-like feature of a Cooper pair on the 
nuclear surface, as described in Ref. \cite{HSCP07}. 

We also show in the lower panel the local 
expectation value of the $nn$-interaction, $E_{\rm pair}(R)$. 
This quantity is defined as 
%
%
%
\begin{equation}
\langle \Psi_{\rm gs}|v_{\rm pair}|\Psi_{\rm gs}\rangle 
=\int\,d\vec{R}d\vec{r}\,|\Psi_{\rm gs}(\vec{R},\vec{r})|^2
v_{\rm pair}(\vec{r}_1,\vec{r}_2) 
\equiv\int R^2dR\,E_{\rm pair}(R), 
\end{equation}
where $\vec{r}_1$ and $\vec{r}_2$ are the coodinates of the valence 
neutrons relative to the core nucleus, and $\vec{r}= \vec{r}_1-\vec{r}_2$ 
and $\vec{R}=(\vec{r}_1+\vec{r}_2)/2$. The ground state wave function 
is denoted as $\Psi_{\rm gs}$. 
We see that 
there is maximal attraction where the 
the Cooper pair is smallest ($R\sim$ 3 fm).
Note that this surface enhancement of the $nn$ force is {\it not} 
primarily an effect of the density effect because it also happens 
with the density independent Gogny force\cite{PSS07}. 
It namely is just the other way round: the density dependence 
of a zero range force is needed to mock up the {\it finite} 
range of a {\it density independent} force, as is explained 
in Ref. \cite{GSMS99}. 
In fact, the similar compact Cooper pair on the nuclear surface 
has recently been shown in Ref.
\cite{PSS07} using the
finite range
Gogny interaction as well as in an
old study with a
simple Yukawa interaction\cite{I77}.

\begin{figure}[htb]
\centerline{\psfig{file=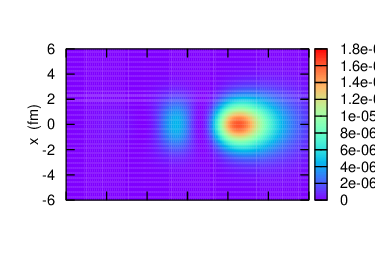,width=7cm,clip}}
\vspace{-1.7cm}
\centerline{\psfig{file=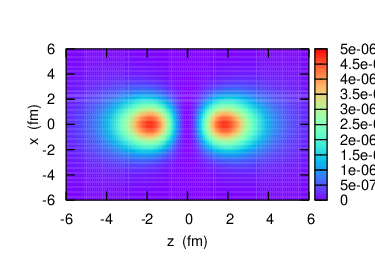,width=7cm,clip}}
\caption{(Color online)
A two-dimensional (2D) plot for
the two-particle density for 
the correlated pair (the upper panel) and for the 
uncorrelated [(1p$_{1/2})^2$] configuration (the lower panel). 
It represents the probability distribution for the spin-up neutron
when the spin-down neutron is at $(z,x)=(3.4,0)$ fm. 
}
\end{figure}
 
The dramatic influence of the pairing interaction can be clearly seen
also in the two-particle density. 
Figure 2 shows the two-particle density for the 
correlated and the uncorrelated 
[(1p$_{1/2})^2$] 
configurations 
in the total spin $S=0$ channel when the spin-down
neutron is located at $(z,x)=(3.4,0)$ fm. 
As can be seen, in the non-interacting case, the 
distribution has a symmetric two bump structure with respect 
to the origin of the nucleus. 
This originates from the absence of 
mixing of wave functions of opposite parity. 
On the contrary, in 
the interacting case, the bump on the far side almost disappears 
and only the bump on the side of the test particle survives. This is
the essential effect of major shell mixing, which is a clear indication of 
{\it strong pairing correlations} \cite{PSS07}. 

\section{Di-neutron structure in $^8$He}

Let us now discuss the di-neutron structure in the $^8$He nucleus. 
An important question is
how the spatial structure of valence neutrons
evolves
from that in the 2$n$-halo nucleus, $^{11}$Li,
when there are more numbers of neutrons.
To address this, we study the  $^8$He nucleus using a core+4$n$ 
five-body model \cite{HTS08}. 
We again use the density-dependent contact pairing interaction among 
the valence neutrons, and diagonalize the five-body Hamiltonian 
with the Hartree-Fock-Bogoliubov (HFB) + particle number projection method. 
See Ref. \cite{HTS08} for further details. 

\begin{figure}[htb]
\centerline{\psfig{file=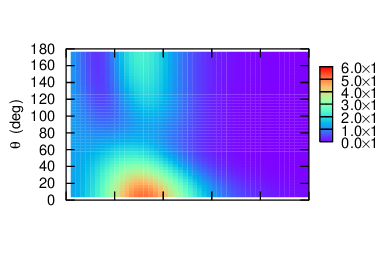,width=7cm}}
\vspace{-1.7cm}
\centerline{\psfig{file=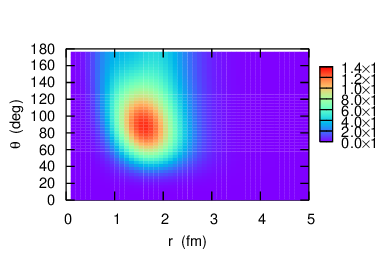,width=7cm}}
\caption{
The two-particle density for
the $^8$He nucleus as a function of $r_1=r_2=r$ and the relative angle
$\theta$ between a spin-up and a spin-down neutrons (the upper panel).
The lower panel is for the
four-particle
density for the dineutron-dineutron configuration.
}
\end{figure}
 
The top panel in Fig. 3 shows the two-particle
density,
$\rho_2(r,\hat{\vec{r}}=0,\uparrow;r,\hat{\vec{r}},\downarrow)$. 
One clearly finds a strong concentration of
two-particle density around $\theta\sim 0$ at around the nuclear
surface. This is similar to what has been found in the Borromean
nucleus $^{11}$Li, and clearly indicates the strong di-neutron correlation
in this nucleus. 
Since the strong di-neutron structure is apparent for a spin-up and 
spin-down neutrons, 
we next plot in the lower panel 
the four-particle density for the two-dineutron configuration.  
The four-particle density for the
dineutron-dineutron configuration has a peak around $\theta\sim \pi/2$.
This peak arises from the main component of the wave
function, that is, the [(1p$_{3/2})^4$] configuration, for which the
four-particle density is proportional to $\sin^4\theta\propto
|Y_{11}|^4$. 
Therefore, two di-neutrons seem to move rather freely in the core+4n nuclei
respecting solely the Pauli principle.

\section{Conclusion}

We studied the two-neutron wave function
in the Borromean nucleus $^{11}$Li by using a three-body model with
a density-dependent pairing force,
and explored the spatial distribution of
the two neutron wave
function as a function of
the center of mass distance $R$ from the core nucleus. 
We showed that the relative distance
between the two neutrons
has a pronounced minimum on the nuclear surface, that is, a {\it compact}  
BEC-like di-neutron structure. 
We also studied the di-neutron structure in the $^8$He nucleus. 
We showed that 
two neutrons with the coupled spin of $S$=0
exhibit a strong di-neutron correlation around the surface of
this nucleus, whereas
the correlation between the two di-neutrons is much weaker.

\section*{Acknowledgements}

This work was supported by the Japanese
Ministry of Education, Culture, Sports, Science and Technology
by Grant-in-Aid for Scientific Research under
the program numbers 19740115 and C 20540277.

\end{document}